\documentclass[letterpaper, 10 pt, conference]{ieeeconf}  

\IEEEoverridecommandlockouts                              
\usepackage{amsmath}
\usepackage{amsfonts}
\usepackage{amssymb}
\usepackage{theorem,pifont}
\usepackage{pstricks}
\usepackage{graphicx}
\usepackage{psfrag,color}
\usepackage{pst-all}
\usepackage{stmaryrd}

\newcommand{\R}{\mathbb{R}}             

\title{\LARGE \bf
For time-invariant delay systems, global asymptotic stability \newline does not imply uniform global attractivity
}

\author{Antoine Chaillet, Fabian Wirth, Andrii Mironchenko and Lucas Brivadis
\thanks{This work has been supported by BayFrance, project  FK-20-2022. 
}
\thanks{A. Chaillet and L. Brivadis are with Universit\'e Paris-Saclay, CNRS, CentraleSup\'elec, Laboratoire des signaux et syst\`emes, 91190, Gif-sur-Yvette, France.
        {\tt\footnotesize antoine.chaillet, lucas.brivadis@centralesupelec.fr}}%
\thanks{F. Wirth is with the Faculty of Computer Science and Mathematics, University of Passau, 94032 Passau, Germany.
        {\tt\footnotesize fabian. (lastname)@uni-passau.de}}
\thanks{A. Mironchenko is with the Department of Mathematics, University of Klagenfurt, 9020, Klagenfurt, Austria. 
{\tt\small andrii.mironchenko@aau.at}
}
}

\newtheorem{proposition}{Proposition}
\newtheorem{lemma}{Lemma}
 
\newtheorem{definition}{Definition}

\begin{document}

\maketitle
\thispagestyle{empty}
\pagestyle{empty}

\begin{abstract}
Adapting a counter-example recently proposed by J.L. Mancilla-Aguilar and H. Haimovich, we show here that, for time-delay systems, global asymptotic stability does not ensure that solutions converge uniformly to zero over bounded sets of initial states. Hence, the convergence might be arbitrarily slow even if initial states are confined to a bounded set. 
\end{abstract}






\section{Introduction}

While ubiquitous in physics, engineering, biology or economics \cite{Niculescu01,Fridman:2014vo,Gu:2012vm}, time-delay systems (TDS) are be more challenging to analyze than ordinary differential equation (ODE) models. This mathematical complexity arises from the infinite-dimensional nature of the state, which is no longer a vector of a Euclidean space, but rather a signal defined over an interval of length equal to the maximal delay involved.

As recently reviewed in \cite{chaillet2023iss}, despite this infinite-dimensionality, a lot of tools valid for ODE models, including Lyapunov analysis, extend rather naturally to TDS. Yet, some fundamental differences remain. For instance, it was not clear until very recently whether the combination of global attractivity of an equilibrium combined with its Lyapunov stability (meaning global asymptotic stability (GAS)) is enough to guarantee uniform global asymptotic stability (UGAS), which additionally imposes that the rate at which solutions converge and their maximal transient overshoot are uniform over bounded sets of initial states \cite{karafyllis2022global}. 

Similarly, it was not known until recently whether the existence of solutions at all positive times (forward completeness (FC)) ensures that the trajectories starting from a bounded set remain in a bounded set on any bounded time interval. This property is referred to in the literature as the bounded reachability set property (BRS, \cite{mironchenko2020input,Mironchenko:2017wb}) or as robust forward completeness (RFC, \cite{KAJIbook11,karafyllis2022global}).



The answer to the two above questions is positive for finite-dimensional systems. 
Namely, for systems described by an ODE with locally Lipschitz right-hand side, GAS is equivalent to UGAS \cite{ROUHABLAL,YOSH} and FC is equivalent to BRS \cite{LINSONWAN}.
These features turned out to be instrumental in the development of stability theory for ODE systems, particularly for converse Lyapunov results \cite{teel2000smooth,LINSONWAN,ANGSON-UFC}.



In contrast, it was shown in \cite[Example 2]{MiW18b} that these equivalences do not extend to infinite-dimensional systems. More recently, the example given in \cite{mancilla2023time} shows that these equivalences do not hold even when focusing on TDS. More precisely, in \cite{mancilla2023time}, a TDS with three state variables and a single discrete delay is given, which is GAS (hence, FC) and whose origin is locally exponentially stable but which is neither UGAS nor even BRS. The solutions of that system may exhibit arbitrarily large transients even when initial states are constrained to the unit ball.

Surprisingly, despite these arbitrarily large transients, the origin of the example in \cite{mancilla2023time} remains uniformly globally attractive (UGA), meaning that the time needed to reach a given neighborhood of the origin is uniform over bounded sets of initial states. Hence, it is not clear yet whether GAS implies UGA. This question is not motivated by mathematical curiosity only, but is also of practical significance as a positive answer would ensure that a TDS cannot have arbitrarily slow transients for initial states in a bounded set.



In this paper, we give a negative answer to this question. By modifying the example in \cite{mancilla2023time}, we show that GAS does not imply UGA for TDS, even when combined with local exponential stability of the origin. We even show that the time needed to first touch a given neighborhood of the origin can be arbitrarily large even from a bounded set of initial conditions. The proposed example is made of four state variables and two discrete delays, which shows that this lack of relation between GAS and UGA is not a feature of a particularly weird class of TDS.

After introducing the necessary background and definitions in Section \ref{sec-prel}, we present the counter-example in Section \ref{sec-main}. 
All proofs are provided in Section \ref{sec-proofs}.

\section{Preliminaries}\label{sec-prel}

\subsection{Notation}
Given $x\in\mathbb R^n$, $|x|$ denotes its Euclidean norm and $|A|$ denotes the induced matrix norm of $A\in \R^{n\times n}$. Given intervals $\mathcal I,\mathcal J\subset\mathbb R$, $C(\mathcal I,\mathcal J)$ denotes the set of continuous functions from $\mathcal I$ to $\mathcal J$. Given $\theta>0$, $\mathcal X:=C([-\theta,0],\mathbb R)$. $\mathcal U$ denotes the set of all signals $u:\mathbb R_{\geq 0}\to\mathbb R$ that are Lebesgue measurable and locally essentially bounded. Given an interval $\mathcal I\subset \mathbb R_{\geq 0}$ and a locally essentially bounded signal $u:\mathcal I\to\mathbb R^m$, $\|u\|:=\textrm{ess\,sup}_{t\in\mathcal I} |u(t)|$. Given $u\in\mathcal U^m$, $u_{\mathcal I}:\mathcal I\to\mathbb R^m$ denotes its restriction to the interval $\mathcal I$, in particular $\|u_{\mathcal I}\|=\textrm{ess\,sup}_{t\in\mathcal I} |u(t)|$. Given $T\in \mathbb R_{\geq 0}\cup\{+\infty\}$, $\theta>0$, $x\in C([-\theta,T),\mathbb R^n)$ and $t\in [0,T)$, $x_t\in\mathcal X^n$ is the history function defined as $x_t(\tau):=x(t+\tau)$ for all $\tau\in[-\theta,0]$.


\subsection{Definitions}

Consider a time-delay system defined by
\begin{align}\label{eq-sys}
    \dot x(t) = f(x_t),
\end{align}
where $f:\mathcal X^n\to\mathbb R^n$ is assumed to be Lipschitz on bounded sets and to satisfy $f(0)=0$. Given $x_0\in\mathcal X^n$, we denote by $x(\cdot;x_0): \mathcal{I}_{\max}(x_0)\to \mathbb{R}^n$ (or simply $x(\cdot)$ when the initial state is clear from the context) the maximal solution of \eqref{eq-sys}. We recall below some classical notions \cite{chaillet2023iss}.


\begin{definition}[FC, BRS]
    The system \eqref{eq-sys} is said to be \emph{forward complete (FC)} if for all initial states $x_0\in\mathcal X^n$ the corresponding solution exists for all positive times. 
    
    It is said to have \emph{bounded reachability sets (be BRS)} if, in addition, given any $\Delta,T>0$, there exists $R>0$ such that, for all $x_0\in\mathcal X^n$ with $\|x_0\|\leq \Delta$, the corresponding solution satisfies $|x(t;x_0)|\leq R$ for all $t\in[0,T]$.
\end{definition}

FC imposes the existence of all solutions at all positive times, which means that no solution can explode in finite time: see \cite[Theorem 2]{chaillet2023iss} or \cite[Theorem 3.2, p. 43]{Hale:1977tr}. BRS additionally imposes that, over bounded time intervals, any solution starting with an initial state in a prescribed ball of radius $\Delta$ remains inside a ball of radius $R$. It is worth stressing that BRS is sometimes referred to as robust forward completeness (RFC) in the literature \cite{KAJIbook11,karafyllis2022global,chaillet2023iss,mancilla2023time}. 



It was recently proved in \cite{mancilla2023time} that, unlike for ODE systems \cite{LINSONWAN}, FC does not imply BRS for time-delay systems.

The next definition characterizes attractivity of the origin\footnote{Note that we tacitly assume that the system \eqref{eq-sys} is FC in Definitions \ref{defGA}-\ref{defLES}.}.

\begin{definition}[GA, UGA]\label{defGA}
    The origin of \eqref{eq-sys} is called \emph{globally attractive (GA)} if $\lim_{t\to\infty} x(t;x_0)=0$ for all $x_0\in\mathcal X^n$. 
    
    It is called \emph{uniformly globally attractive (UGA)} if, given any $\varepsilon,\Delta>0$, there exists $T\geq 0$ such that, for all $x_0\in\mathcal X^n$ with $\|x_0\|\leq \Delta$, it holds that $|x(t;x_0)|\leq \varepsilon$ for all $t\geq T$.
\end{definition}

Both GA and UGA impose that all solutions of \eqref{eq-sys} asymptotically converge to the origin. In UGA, the convergence rate is additionally uniform over bounded sets of initial states.

We may weaken these properties by considering the case when solutions approach any arbitrarily small neighborhood of the origin, without being necessarily trapped in it afterwards. This leads to the notion of weak attractivity \cite{MiW19a}.

\begin{definition}[WGA, WUGA]
    The origin of \eqref{eq-sys} is called \emph{weakly globally attractive (WGA)} if $\inf_{t\geq 0} |x(t;x_0)|=0$ for all $x_0\in\mathcal X^n$. 

    It is called \emph{weakly uniformly globally attractive (WUGA)} if, for all $\varepsilon,\Delta>0$, there exists $T\geq 0$ such that, for all $x_0\in\mathcal X^n$ with $\|x_0\|\leq \Delta$, $|x(t;x_0)|\leq \varepsilon$ for some $t\in[0,T]$.
\end{definition}

The key difference with the two lies in the fact that, for WUGA, the time needed to touch a given neighborhood of the origin is uniform over bounded sets of initial states.

When global attractivity is combined with stability, we get the following concepts.

\begin{definition}[GAS, UGAS] The system \eqref{eq-sys} is said to be \emph{globally asymptotically stable (GAS)} if its origin is stable (in the sense that, given any $\varepsilon>0$, there exists $\delta>0$ such that, for all $x_0\in\mathcal X^n$ with $\|x_0\|\leq \delta$ it holds that $|x(t;x_0)|\leq \varepsilon$ for all $t\geq 0$) and globally attractive.

The system \eqref{eq-sys} is said to be \emph{uniformly globally asymptotically stable (UGAS)} if there exists $\beta\in\mathcal{KL}$ such that, for all $x_0\in\mathcal X^n$, it holds that $|x(t;x_0)|\leq \beta(\|x_0\|,t)$ for all $t\geq 0$.
\end{definition}

While GAS guarantees Lyapunov stability and convergence of all solutions to the origin, the $\mathcal{KL}$ state estimate given by UGAS additionally imposes that both the convergence rate and the size of transient overshoots be uniform over bounded sets of initial states. In other words, for a UGAS system, it is not possible to have arbitrarily slow convergence or arbitrarily large transients when considering a bounded set of initial states, which constitutes an interesting feature in practice. But the interest of this $\mathcal{KL}$ estimate goes beyond this. In particular, it was instrumental in deriving converse Lyapunov results for ODE systems \cite{teel2000smooth,LINSONWAN}.
    
Here again, the example proposed in \cite{mancilla2023time} shows that, unlike for ODE systems \cite{ROUHABLAL,YOSH}, GAS does not necessarily guarantee UGAS for TDS.

We finally recall the notion of exponential stability. 

\begin{definition}[LES]\label{defLES}
    The origin of \eqref{eq-sys} is said to be \emph{locally exponentially stable (LES)} if there exist $\Delta, k,\mu>0$ such that for all $x_0\in\mathcal X^n$ with $\|x_0\|\leq \Delta$ the solution satisfies $|x(t;x_0)|\leq k\|x_0\|e^{-\mu t}$ for all $t\geq 0$.
\end{definition}

\section{The counter-example}\label{sec-main}

\subsection{Main statement}
In \cite{mancilla2023time}, the example of a GAS system that is not UGAS violates a particular feature of UGAS, namely that no arbitrarily large transients can occur from bounded sets of initial states. Yet, the system in \cite{mancilla2023time} surprisingly happens to be UGA, meaning that the convergence rate to the origin is uniform over bounded sets of initial states. Here, we modify that example to show that GAS does not imply UGA, even when combined to LES. Recall that the example in \cite{mancilla2023time} reads
\begin{subequations}\label{eq-haim}
\begin{align}
\dot x(t) =&\,\, g(x(t),z_1(t-\theta)),\label{eq-haima}\\
\dot z_1(t) =&\, -z_1(t),\label{eq-haimb}
\end{align}
\end{subequations}
where $\theta>0$, $x(t)\in\mathbb R^2$, $z_1(t)\in\mathbb R$ and $g:\mathbb R^3\to\mathbb R^2$ is defined for all $x\in\mathbb R^2$ and all $u\in\mathbb R$ as
\begin{align*}
g(x,u):=(1+|x|^2)\big(\varphi(u)A_1+(1-\varphi(u))A_0\big) x,
\end{align*}
with the function $\varphi:\mathbb R\to[0,1]$ defined as
\begin{align*}
\varphi(s):=\left\{\begin{array}{ll}
0 & \textrm{ if } s<0,\\
s & \textrm{ if } s\in[0,1],\\
1 & \textrm{ if } s>1,
\end{array}\right.
\end{align*}
and the matrices
\begin{align*}
A_0:=\begin{pmatrix}
-0.1 & 0.5 \\
-2 & 0
\end{pmatrix},\qquad
A_1:=\begin{pmatrix}
0 & 2 \\
-0.5 & -0.1
\end{pmatrix}.
\end{align*}
Given $c>0$, consider the time-delay system defined as
\begin{subequations}\label{eq-10}
\begin{align}
\dot x(t) =&\,\, c\varphi(z_2(t-2)) g(x(t),z_1(t-1))\nonumber\\
&+c(1-\varphi(z_2(t-2)))A_0x(t),\label{eq-10a}\\
\dot z_1(t) =&\, -z_1(t),\label{eq-10b}\\
\dot z_2(t) =&\,-z_2(t).\label{eq-10c} 
\end{align}
\end{subequations}
As the system has two delays ($1$ and $2$), we now consider $\mathcal{X}:= C([-2,0],\mathbb{R})$.
We indicate by $X:=(x,z_1,z_2)^\top\in\mathcal X^4$ the state \eqref{eq-10} and by $X_0:=(x_0,z_{10},z_{20})^\top\in\mathcal X^4$ the corresponding initial state. Accordingly, we denote by $X(\cdot;X_0):\mathcal{I}_{\max}(X_0) \to \mathbb{R}^4$ (or simply $X(\cdot)$) its solution starting from $X_0\in \mathcal X^4$ over its maximal interval of existence.

As compared to \eqref{eq-haim}, this example adds the extra dynamics $z_2$. It also picks the delay $\theta=1$ in \eqref{eq-haima} but adds a further delay term to the system, so that the overall delay is $2$.
Also there is an extra parameter $c>0$, which will be used to adapt the evolution speed of the $x$-solutions. We establish the following in Sections \ref{sec-proof1} to \ref{sec-proof6}.

\begin{proposition}[GAS \& LES $\nRightarrow$ WUGA]\label{prop-main}
Given any $c>0$, the time-delay system \eqref{eq-10} 
\begin{enumerate}
\item is FC, \label{item0}
\item is LES, \label{item2}
\item is GAS. \label{item4}
\end{enumerate}   
However, there exists $c>0$ such that the TDS \eqref{eq-10}
\begin{enumerate}
 \setcounter{enumi}{3}
\item is not BRS, \label{item1}
\item is not UGA: more precisely, given any $T>0$, there exists an initial state $X_0\in\mathcal X^4$ with $\|X_0\|\leq 2$ such that the solution of \eqref{eq-10} satisfies $|X(T;X_0)|\geq 1$,\label{item5}
\item is not WUGA. \label{item6}
\end{enumerate}   
\end{proposition}


\subsection{Rationale behind this example}

We informally present the rationale behind this result. 

In view of \eqref{eq-10b}-\eqref{eq-10c}, both $z_1$ and $z_2$ converge exponentially fast to zero and remain small at all times if they are small at time $t=0$. In particular, for small enough initial states, \eqref{eq-10a} essentially behaves as $\dot x(t) = A_0x(t)$ and exponential convergence of $x$ follows from the fact that $A_0$ is Hurwitz. This explains intuitively why \eqref{eq-10} is LES. 

For GAS, since we already have Lyapunov stability of the origin, it is sufficient to show that all solutions eventually vanish. This is done by observing that, since both $z_1$ and $z_2$ tend to zero, \eqref{eq-10a} behaves asymptotically like $\dot x(t) = A_0x(t)$, thus showing that $x$ also converges to zero. 

The fact that \eqref{eq-10} is FC but not BRS was the main contribution of \cite{mancilla2023time}. It exploits the fact that, if $z_1$ were able to switch discontinuously between $0$ and $1$, then \eqref{eq-haima} would exhibit finite escape times \cite{mancilla2005representation}. For  continuous switching signals $z_1$, finite escape times no longer occur, but the solutions of \eqref{eq-haim} can reach any arbitrarily large state norm values by conveniently choosing the initial state $z_{10}$ in the ball of radius $1$, which contradicts BRS. By adjusting the value of the parameter $c$ and by choosing $z_{20}$ identically equal to $1$ on $[-2,-1]$, we have that \eqref{eq-10a} behaves qualitatively like \eqref{eq-haima} on $[0,1]$ and its state reaches arbitrarily large values over this time interval, thus contradicting BRS. 

The absence of uniform attractivity is the result of the additional $z_2$ dynamics (recall that \eqref{eq-haim} is UGA). The idea is that, at $t=1$, $|x(t)|$ has reached an arbitrarily large value. By choosing $z_{20}$ conveniently over the interval $[-1,0]$, we can show that the convergence of $x$ to zero is at most exponential. So the time needed to reach any given neighborhood of the origin is arbitrarily long, which is incompatible with UGA. The lack of WUGA follows along the same ideas, but is formally established by recalling that if \eqref{eq-10} were WUGA, then stability (Item \ref{item2}) would guarantee UGA.










The proofs of Items \ref{item1}) and \ref{item5}) rely on the following observation, which is established in Section \ref{proof-lemma1}. It was shown in \cite{mancilla2023time} that the system \eqref{eq-haim} is not BRS. The proof employed there consisted in seeing $z_{1}$ as a generic input and in showing that the corresponding system with input is not forward complete by relying on \cite[Example 3.5]{mancilla2005representation}. The conclusion followed implicitly by showing that such forward completeness was a necessary and sufficient condition for the BRS of the corresponding TDS. Here, we provide a more constructive proof by building a solution whose norm overpasses any prescribed value over a bounded time interval. 

\begin{lemma}[Arbitrarily large transients]\label{lemma1}
There exists a constant $c>0$ such that, given any $M>0$, there exists an initial condition $z_{10}\in\mathcal X$ satisfying
\begin{align*}
\|z_{10}\|\leq 1,\quad z_{10}(0)=0
\end{align*}
and some $x_0\in\mathcal X^2$ with $\|x_0\|\leq 1$ such that, given any $z_{20}\in\mathcal X$ satisfying $z_{20}(t)=1$ for all $t\in[-2,-1]$, the corresponding solution of \eqref{eq-10} satisfies
\begin{align*}
    |x(1)|\geq 2M.
\end{align*}
\end{lemma}

Not that the absence of BRS is in line with \cite[Theorem 1]{karafyllis2022global}, where it is shown that GAS \& BRS is equivalent to UGAS.


\section{Proofs}\label{sec-proofs}

\subsection{Proof of Lemma \ref{lemma1}}\label{proof-lemma1}

\begin{figure*}
    \centering
    \includegraphics[width=0.9\textwidth]{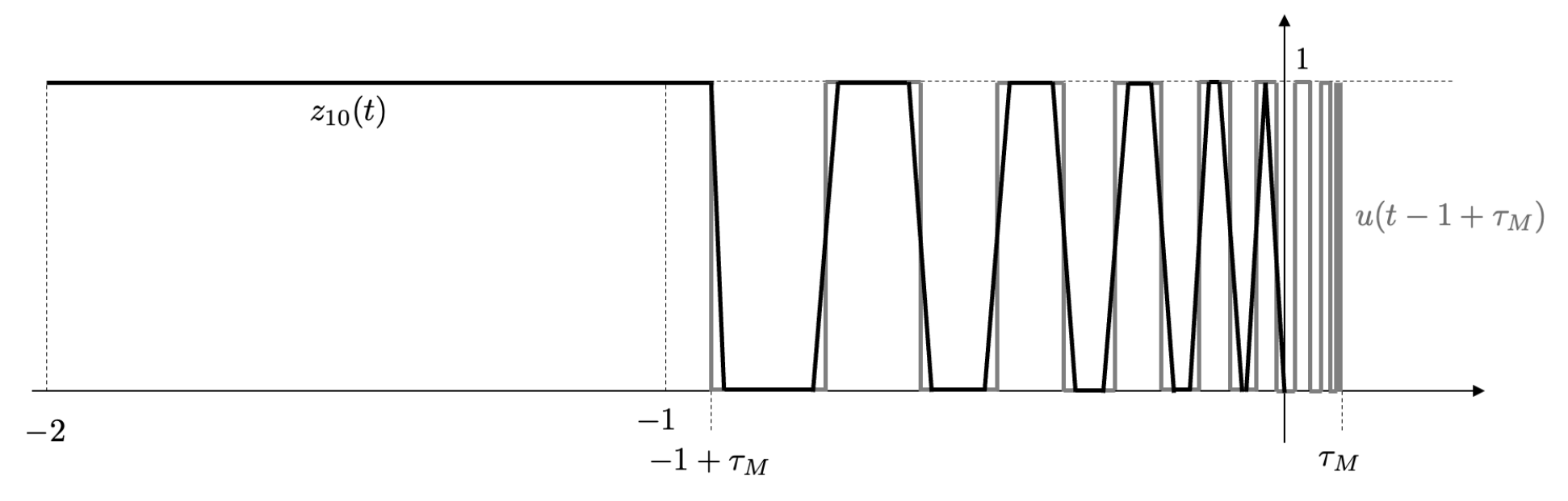}
    \caption{Graphical representation of $z_{10}$.}
    \label{fig:z10}
\end{figure*}

Consider any $c>0$ and any initial state $X_0=(x_0,z_{10},z_{20})^\top\in\mathcal X^4$ as in the statement, namely with $z_{20}(t)=1$ for all $t\in[-2,-1]$. The solution of \eqref{eq-10} then satisfies, for 
all $t\in[0,1]$,
\begin{align*}
    \dot x(t)=&\,\,c(1+|x(t)|^2)\Big(\varphi(z_1(t-1))A_1\\
    &+(1-\varphi(z_1(t-1))A_0\Big)x(t),
\end{align*}
which coincides with the dynamics of \eqref{eq-haim} when $c=1$. Consider the corresponding ODE system with inputs:
\begin{align}\label{eq-21}
    \dot w=c(1+|w|^2)\left(uA_1+(1-u)A_0\right)w.
\end{align}
Given $u\in\mathcal U$ and $w_0\in\mathbb R^2$, denote by $w(\cdot;w_0,u)$ the corresponding solution over its maximal interval of existence. It was shown in \cite[Example 3.5]{mancilla2005representation}
that system \eqref{eq-21} with $c=1$ is not forward complete. In particular, for $w_0=(0, 1/2)^\top$, there exists a piecewise constant signal $u:\mathbb R_{\geq 0}\to\{0,1\}$ and a time $T>0$ such that the corresponding solution of \eqref{eq-21} with $c=1$ satisfies $\lim_{t\to T^-}|w(t;w_0,u)|=+\infty$. As a consequence, by picking $c=T$, the solution of \eqref{eq-21} satisfies 
\begin{align}\label{eq-20}
    \lim_{t\to 1^-}|w(t;w_0,u)| =+\infty.
\end{align}
As the vector field in \eqref{eq-21} is locally Lipschitz, its solutions exist in backward time. More precisely, extending the signal $u$ over $(-\infty,0)$ by letting $u(t)=1$ for all $t< 0$, there exists $\bar\tau>0$ such that $w(\cdot;w_0,u)$ exists on $[-\bar\tau,1)$. With no loss of generality, we can pick $\bar\tau\in(0,1)$ small enough that $|w(t)|\leq 1$ for all $t\in[-\bar\tau,0]$. Due to \eqref{eq-20}, given any $M>0$, there exists $\tau_M\in(0,\bar\tau]$ such that 
\begin{align*}
    |w(1-\tau_M;x_0,u)|\geq 3M.
\end{align*}
We now proceed to providing a continuous approximation of this piecewise-constant signal $u$. Over the interval $[-\bar\tau,1-\tau_M]$, the solution $w(\cdot;w_0,u)$ is bounded. As the system \eqref{eq-21} is affine in the input, its flow has continuous dependence with respect to $u$ in the weak-$*$ topology over $[-\bar\tau,1-\tau_M]$: see \cite[Theorem 3.1]{gauthier2001deterministic}. Pick any sequence\footnote{Recall that $C([-\bar\tau,1-\tau_M],\R)$ is weak-$*$ dense in $L^\infty([-\bar\tau,1-\tau_M],\R)$.} $\{u_k\}_{k\in\mathbb N}\subset C([-\bar\tau,1-\tau_M],[0,1])$ that converges to $u_{[-\bar\tau,1-\tau_M]}$ in the weak-$*$ topology and that satisfies $u_k(-\bar\tau)=1$ and $u_k(1-\tau_M)=0$. By continuous dependence, we then have that $\lim_{k\to+\infty} w(1-\tau_M;w_0,u_k)=w(1-\tau_M;w_0,u)$. Extend $u_k$ to $[-\bar\tau,+\infty)$ by letting $u_k(t)=1$ for all $t\in[-\bar\tau,0]$ and $u_k(t)=0$ for all $t\geq 1-\tau_M$. We conclude that for some $K\in\mathbb N$ large enough, 
\begin{align}\label{eq-22}
    |w(1-\tau_M;w_0,u_K)|\geq 2M. 
\end{align}
Now, consider the initial state $X_0=(x_0,z_{10},z_{20})^\top\in \mathcal X^4$ where $x_0(t):=w(-\tau_M;w_0,u_K)$ for all $t\in[-2,0]$, $z_{20}\in\mathcal X$ is any signal satisfying $z_{20}(t)=1$ for all $t\in[-2,1]$, and
\begin{align*}
    z_{10}(t):=\left\{\begin{array}{cl}
       1  & \textrm{ if } t\in[-2,-1)\\
       u_K(t+1-\tau_M)  & \textrm{ if } t\in [-1,0],
    \end{array}\right.
\end{align*}
as depicted by Figure \ref{fig:z10}.
Then, as claimed, $\|z_{10}\|=1$ and $z_{10}(0)=u_K(1-\tau_M)=0$. Furthermore, the corresponding solution of \eqref{eq-10} satisfies
\begin{align*}
    \dot x(t)=&\,\,c(1+|x(t)|^2)u_K(t-\tau_M)A_1x(t)\\
    &+(1-u_K(t-\tau_M))A_0x(t),\quad \forall t\in [0,1].
\end{align*}
Consequently, $x(t)=w(t;x_0(0),u_K(\cdot-\tau_M))=w(t;w(-\tau_M),u_K(\cdot-\tau_M))$ for all $t\in [0,1]$, where $w(-\tau_M):=w(-\tau_M;w_0,u_K(\cdot-\tau_M))$. By the cocycle property, it holds that $w(t;w(-\tau_M),u_K(\cdot-\tau_M))=w(t-\tau_M;w_0,u_K)$. We conclude from \eqref{eq-22} that $|x(1)|\geq 2M$.

The following subsections present the proof for the different statements of Proposition~\ref{prop-main}.

\subsection{Proof of \ref{item0}): forward completeness}\label{sec-proof1}

The proof of FC uses the same arguments as those in \cite{mancilla2023time}: we report it here for completeness. Consider any $c>0$. In view of \eqref{eq-10b}-\eqref{eq-10c}, given any $z_{10},z_{20}\in\mathcal X$, $z_1(\cdot)$ and $z_2(\cdot)$ exist at all positive times and are continuous on $[-2,+\infty)$. Consequently, it is sufficient to show that the system
\begin{align}
\dot x &= c\varphi(u_2) g(x,u_1) +c(1-\varphi(u_2))A_0x \label{eq-11}
\end{align}
is forward complete with respect to continuous inputs, meaning that its solutions exist at all positive times when $u:=(u_1,u_2)^\top\in C(\mathbb R_{\geq 0},\mathbb R^2)$. We proceed by contradiction: assume to the contrary that there exists an input $u\in C(\mathbb R_{\geq 0},\mathbb R^2)$, an initial state $x_0\in\mathbb R^2$ and a time $T>0$ such that the solution $x(\cdot)$ of \eqref{eq-11} exists on $[0,T)$ and satisfies
\begin{align}\label{eq-12}
\lim_{t\to T^-} |x(t)|=+\infty.
\end{align}
Consider the matrices defined as
\begin{align}\label{eq-28}
    A_\lambda:=\lambda A_1+(1-\lambda)A_0,\quad  \forall\lambda\in[0,1].
\end{align}
Each $A_\lambda$ being Hurwitz \cite[Lemma 6]{mancilla2023time}, there exists a symmetric positive definite matrix $P_\lambda\in\mathbb R^{2\times 2}$ such that 
\begin{align}\label{eq-24}
A_\lambda^\top P_\lambda+P_\lambda A_\lambda^\top =-I.
\end{align}
For short, let $\bar\varphi_2(t):=\varphi(u_2(t))$,  $A(t):=A_{\bar \varphi_2(t)}$ and $P(t):=P_{\bar \varphi_2(t)}$ for all $t\geq 0$. Then we have in particular that $A(T)^\top P(T)+P(T)A(T)=-I$. Since $\bar\varphi_2$ is continuous, there exists $\Delta>0$ such that, for all $t\in[T-\Delta,T]$,
\begin{align}\label{eq-23}
A(t)^\top P(T)+P(T)A(t) \leq -\frac{1}{2}I.
\end{align}
Considering the function $V:\mathbb R^2\to\mathbb R_{\geq 0}$ defined as $V(x)=x^\top P(T) x$ for all $x\in\mathbb R^2$, it follows that, along the solutions of \eqref{eq-11} and for all $t\in[T-\Delta,T)$,
\begin{align*}
\dot V =&\,\, 2x(t)^\top P(T)\Big( c\bar\varphi_2(t)A_0x(t)\\
&\,\,
+c(1-\bar\varphi_2(t))(1+|x(t)|^2)A(t)x(t)\Big)\\
=&\,\, c(1-\bar\varphi_2(t))(1+|x(t)|^2)x(t)^\top \big(P(T)A(t)\\
&\,\,+A(t)^\top P(T)\big)x(t)+2c\bar\varphi_2(t)x(t)^\top P(T)A_0x(t)\\
\leq & -\frac{c}{2}(1-\bar\varphi_2(t))(1+|x(t)|^2)|x(t)|^2\\
&\,\,+2c|P(T)A_0||x(t)|^2.
\end{align*}
As $V$ is quadratic and positive definite, there exists $\tilde c>0$ such that $2c|P(T)A_0||x|^2\leq \tilde cV(x)$. Since $\bar\varphi_2(t)\leq 1$, we conclude that $\dot V\leq \tilde cV(x(t))$ for all $t\in[T-\Delta,T)$, hence
\[
\limsup_{t\to T^-}V(x(t))\leq V(x(T-\Delta))e^{\tilde c\Delta}<+\infty,
\]
in contradiction to \eqref{eq-12}. Thus, the system \eqref{eq-10} is FC.

\subsection{Proof of \ref{item2}): local exponential stability}\label{sec-proof-fact-LES}
Here also, the proof is similar to the one given in \cite{mancilla2023time}. By continuity, we get from \eqref{eq-24} that there exists $\bar\lambda\in(0,1)$ such that the matrix $A_\lambda$ defined in \eqref{eq-28} satisfies
\begin{align}\label{eq-25}
    A_\lambda^\top P_0+P_0A_\lambda=-\frac{1}{2}I,\quad \forall\lambda\in[0,\bar\lambda].
\end{align}
Observe that, from \eqref{eq-10b}-\eqref{eq-10c}, it holds for all $i\in\{1,2\}$ and all $z_{i0}\in\mathcal X$ that
\begin{align}\label{eq-27}
    |z_i(t)|=|z_{i0}(0)|e^{-t}\leq \|z_{i0}\|e^{-t},\quad \forall t\geq 0.
\end{align}
Consider any $c>0$ and any initial state $X_0=(x_0,z_{10},z_{20})^\top \in \mathcal X^4$ with $\|X_0\|\leq \bar\lambda$. Let 
\[
\tilde A(t):=A_{\varphi(z_1(t-1))}=\varphi(z_1(t-1))A_1+\big(1-\varphi(z_1(t-1))\big)A_0
\]
and $\tilde\varphi_2(t):=\varphi(z_2(t-2))$ for all $t\geq 0$. Then \eqref{eq-10a} reads
\begin{align*}
    \dot x(t) = c\tilde \varphi_2(t)(1+|x(t)|^2)\tilde A(t)x(t)+c\big(1-\tilde\varphi_2(t)\big)A_0x(t).
\end{align*}
Since $\tilde\varphi_2(t)\leq \bar \lambda$ for all $t\geq 0$, it holds from \eqref{eq-25} that 
\[
\tilde A(t)^\top P_0+P_0\tilde A(t)\leq -\frac{1}{2}I,\quad \forall t\geq 0. 
\]
Consider the function $V_0$ defined as $V_0(x):=x^\top P_0 x$ for all $x\in\mathbb R^2$. Then there exist $\underline\alpha_0,\overline\alpha_0>0$ such that
\begin{align}\label{eq-26}
    \underline\alpha_0|x|^2\leq V_0(x)\leq \overline\alpha_0|x|^2,\quad \forall x\in\mathbb R^2.
\end{align}
By Item \ref{item0}), the function $v_0(\cdot):=V_0(x(\cdot))$ is well defined on $\mathbb R_{\geq 0}$ and satisfies, for almost all $t\geq 0$,
\begin{align}
\label{eq:v0-dynamics}
    \dot v_0(t)=&\,\, c\tilde\varphi_2(t)(1+|x(t)|^2)x(t)^\top\left(\tilde A(t)^\top P_0+P_0\tilde A(t)\right)x(t)\nonumber\\
    &+\,\, c(1-\tilde\varphi_2(t))x(t)^\top\left(A_0^\top P_0+P_0A_0\right) x(t)\nonumber\\
    \leq&\,\, -\frac{c}{2}\tilde\varphi_2(t)(1+|x(t)|^2)|x(t)|^2-c(1-\tilde \varphi_2(t))|x(t)|^2\nonumber\\
    \leq&\,\,-c\left(1-\tilde\varphi_2(t)/2\right)|x(t)|^2\nonumber\\
    \leq&\,\,-\frac{c}{2\overline\alpha_0}v_0(t).
\end{align}
Consequently, $v_0(t)\leq v_0(0)e^{-\frac{ct}{2\overline\alpha_0}}$, 
which gives with \eqref{eq-26} 
\[
|x(t)|\leq \sqrt{\frac{\overline\alpha_0}{\underline\alpha_0}}\|x_0\|e^{-\frac{ct}{4\overline\alpha_0}}, \quad t\geq 0.
\]
Combining this with \eqref{eq-27}, we conclude that there exist $k,\mu>0$ such that, given any $X_0\in\mathcal X^4$ with $\|X_0\|\leq \bar \lambda$, it holds that $|X(t)|\leq k\|X_0\|e^{-\mu t}$ for all $t\geq 0$. In other words, the origin of \eqref{eq-10} is LES.

\subsection{Proof of \ref{item4}): global asymptotic stability}

Consider any $c>0$. We already know by Item \ref{item0}) that, for every initial condition $X_0\in\mathcal X^4$, the corresponding solution of \eqref{eq-10} exists for all positive times. It is clear that 
\begin{align}\label{eq-29}
    \lim_{t\to+\infty} z_1(t) = \lim_{t\to+\infty} z_2(t) = 0,
\end{align}
 irrespective of initial conditions (see \eqref{eq-27}).  Let $P_0$ be a positive definite symmetric matrix satisfying $A_0^\top P_0 + P_0A_0 = -I$ and choose $\varepsilon>0$ sufficiently small that the matrix defined in \eqref{eq-28} satisfies
\begin{equation}
|z_1| \leq \varepsilon\quad\Rightarrow\quad    A_{\varphi(z_1)}^\top P_0 + P_0 A_{\varphi(z_1)} \leq - \frac{1}{2} I. 
\end{equation}
Fix an initial condition $X_0=(x_0, z_{10},z_{20})^\top \in\mathcal X^4$. 
Let $T>0$ be sufficiently large that $|z_1(t-1)| \leq \varepsilon$ for all $t\geq T$. 
Consider the Lyapunov function  
$V_0(x) = x^\top P_0 x$, and define $v_0(\cdot):= V_0(x(\cdot))$.

Following the arguments in Section \ref{sec-proof-fact-LES},
we have \eqref{eq:v0-dynamics} for almost all $t\geq T$.
This ensures that $\lim_{t\to+\infty} v_0(t)=0$ and consequently that $\lim_{t\to+\infty} x(t)=0$. Considering \eqref{eq-29}, we conclude that the origin of \eqref{eq-10} is globally attractive. Since Item \ref{item2}) ensures Lyapunov stability of the origin, we conclude global asymptotic stability.

\subsection{Proof of \ref{item1}): no boundedness of reachability sets}

Lemma \ref{lemma1} states that there exists a constant $c>0$ such that, given any $M>0$, there exists an initial state $X_0\in\mathcal X^4$ with $\|X_0\|\leq 2$ such that the solution of \eqref{eq-10} satisfies $|X(T)|\geq 2M$. In particular, it holds that $\sup\left\{|X(t)|\,:\, \|X_0\|\leq 2, t\in[0,1] \right\}=+\infty$, which is incompatible with  BRS.

Note that an alternative to the above direct argument is to invoke \cite[Theorem 1]{karafyllis2022global}, which states that UGAS is equivalent to GAS when the BRS property holds. Since the system is GAS  (Item \ref{item4}) and not UGAS (Item \ref{item5}), it cannot be BRS.

\subsection{Proof of \ref{item5}): no uniform global attractivity}

Choose $\lambda_0 >0$ such that $1/2(A_0^\top + A_0) \geq -\lambda_0 I$.
For all $T>1$, we are going to exhibit an initial state $X_0=X_0(T)\in\mathcal X^4$, in the ball of radius $2$, for which convergence to the ball of radius $1/2$ takes longer than the prescribed $T>1$. This contradicts UGA. To that aim, given any $T>1$, consider any $M>0$ satisfying
\begin{align}\label{eq-17b}
    M\geq e^{c\lambda_0T}.
\end{align}
For $\varepsilon\in(0,1]$ define $z_{20}^\varepsilon\in\mathcal X$ by
\begin{align*}
    z_{20}^\varepsilon(t):=\left\{\begin{array}{cl}
       1  & \textrm{ if } t\in[-2,-1] \\
       1-(t+1)/\varepsilon & \textrm{ if } t\in(-1,-1+\varepsilon]\\
        0 & \textrm{ if } t\in[-1+\varepsilon,0].
    \end{array}\right.
\end{align*}
Consider the initial states $x_0\in\mathcal X^2$ and $z_{10}\in\mathcal X$, with $\|x_0\|\leq 1$ and $\|z_{10}\|\leq 1$, and the constant $c>0$ all given by Lemma \ref{lemma1} for this particular value of $M$ and let $X^\varepsilon(\cdot)=(x^\varepsilon(\cdot),z_1(\cdot),z_{2}^\varepsilon(\cdot))^\top$ denote the solution of \eqref{eq-10} with initial state $X_0^\varepsilon:=(x_0,z_{10},z_{20}^\varepsilon)^\top$. Then $\|X_0^\varepsilon\|\leq 2$ and
\begin{align}\label{eq-15b}
    |x^\varepsilon(1)|\geq 2M.
\end{align}
Since $z_{10}(0)=0$, it follows from \eqref{eq-10b} that $z_1(t)=0$ for all $t\geq 0$. Also $\varphi(z_{2}^\varepsilon(t)) = z_{2}^\varepsilon(t)$, $t\geq -2$.
Hence, the solution satisfies for $t\geq 1$:
\begin{align}\label{eq-13b}
    \dot x^\varepsilon(t) = c\left[1+z_2^\varepsilon(t-2)|x^\varepsilon(t)|^2\right]A_0x^\varepsilon(t).
\end{align}
We obtain for $t\geq 1$ along the solutions of \eqref{eq-13b}
\begin{align*}
    \frac{d}{dt} \frac{1}{2}|x^\varepsilon(t)|^2 &=
    c\left[1+z_2^\varepsilon(t-2)|x^\varepsilon(t)|^2\right](x^{\varepsilon}(t))^\top
    A_0x^\varepsilon(t) \\
    &\geq -\lambda_0c\left[1+z_2^\varepsilon(t-2)|x^\varepsilon(t)|^2\right]|x^\varepsilon(t)|^2 \\
    & \geq  -\lambda_0 c\left[1+z_2^1(t-2)|x^\varepsilon(t)|^2\right]|x^\varepsilon(t)|^2.
\end{align*}
As the coefficients of the final differential inequality are independent of 
$\varepsilon\in (0,1]$, it follows from continuity of the solution and a standard comparison result that there is a $\tau\in(0,1)$ such that, for all $\varepsilon\in (0,1]$,
\begin{align}\label{eq-16b}
    |x^\varepsilon(t)|\geq M,\quad \forall t\in [1,1+\tau].
\end{align}
For $\varepsilon\in (0,\tau)$, the system \eqref{eq-13b} simplifies on $[1+\tau,\infty)$ to
$\dot x^\varepsilon(t) = cA_0x^\varepsilon(t)$. On this interval, we then obtain that
\begin{align*}
    \frac{d}{dt} \frac{1}{2}|x^\varepsilon(t)|^2 &\geq  -\lambda_0 c|x^\varepsilon(t)|^2.
\end{align*}
Thus, using \eqref{eq-17b} and \eqref{eq-16b}, 
\begin{align*}
    |x^\varepsilon(T)| &\geq  e^{-\lambda_0 c (T-\tau) } |x^\varepsilon(1+\tau)|\geq e^{\lambda_0c \tau} > 1.
\end{align*}
This completes the proof.

\subsection{Proof of \ref{item6}): no weak uniform global attractivity}\label{sec-proof6}


By (a special case of) \cite[Proposition 4.1]{Mir17a}, the combination of WUGA and ULS implies UGA. 
As our system is ULS but not UGA, we infer that it cannot be WUGA either.

\section{Conclusion}

By extending the counter-example recently proposed by J.L. Mancilla-Aguilar and H. Haimovich, we have shown that, for time-delay systems, UGAS does not ensure uniform global attractivity of the origin, even when combined to local exponential stability. This negative result shows an additional crucial difference with the stability analysis of ODE systems. 

The key point is that, unlike in finite dimension, forward completeness does not guarantee BRS. This example and the original one in \cite{mancilla2023time} plead for a careful assessment of BRS, for which tools exist in the literature: see for instance \cite[Theorem 9]{mancilla2023time} or \cite[Theorems 5 \& 6]{chaillet2023iss}. Alternatively, one may consider using other state-spaces for the study of TDS, in which BRS directly results from forward completeness, such as H\"older or Sobolev spaces \cite{karafyllis2022global} or $L^\infty$ \cite{BCM24}.


\bibliographystyle{plain}
\bibliography{refs.bib}

\end{document}